\documentclass[twocolumn]{aastex62}

\usepackage{float}
\usepackage{graphicx}
\usepackage{dcolumn}
\usepackage{bm}
\usepackage[colorlinks=true]{hyperref}
\usepackage{color}
\usepackage{array}
\usepackage{multirow}
\usepackage[normalem]{ulem}
\usepackage{enumitem,kantlipsum}
\usepackage{soul}

\usepackage[thinc]{esdiff}

\newcommand{\comment}[1]{}

\emergencystretch=\maxdimen
\begin{document}

\title{{\bf A Tale of Two Black Holes: Multiband Gravitational-Wave Measurement of Recoil Kicks}}

\author{Shobhit Ranjan}

\affiliation{Department Physics and Astronomy, Vanderbilt University, 2301 Vanderbilt Place, Nashville, TN, 37235, USA}
\affiliation{Center for Relativistic Astrophysics, Georgia Institute of Technology, Atlanta, GA, 
30332, USA}

\author{Karan Jani}
\affiliation{Department Physics and Astronomy, Vanderbilt University, 2301 Vanderbilt Place, Nashville, TN, 37235, USA}

\author{Alexander H. Nitz}
\affiliation{Department of Physics, Syracuse University, Syracuse, NY 13244, USA}

\author{Kelly Holley-Bockelmann}
\affiliation{Department Physics and Astronomy, Vanderbilt University, 2301 Vanderbilt Place, Nashville, TN, 37235, USA}

\author{Curt Cutler}
\affiliation{Theoretical Astrophysics, California Institute of Technology, Pasadena, CA 91125, USA}
\affiliation{Jet Propulsion Laboratory, California Institute of Technology, 4800 Oak Grove Drive, Pasadena, CA 91109, USA}


\begin{abstract}
The non-linear dynamics of General Relativity leave their imprint on remnants of black hole mergers in the form of a recoil ``kick''. The kick has profound astrophysical implications across the black hole mass range from stellar to super-massive. However, a robust measurement of the kick for generic binaries from gravitational-wave observations has proved so far to be extremely challenging. In this \emph{letter}, we demonstrate the prospects of measuring black hole kicks through a multiband gravitational-wave network consisting of space mission LISA, the current earth-based detector network and a third-generation detector. For two distinct cases of remnant black hole kick (68 km/s, 1006 km/s) emerging from near identical pre-merger configuration of GW190521  -- the first confirmed intermediate-mass black hole -- we find that the multiband network will recover with 90\% credible level the projection of the kick vector relative to the orbital plane within tens of km/s accuracy. Such precise measurement of the kick offer a new set of multi-messenger follow-ups and unprecedented tests of astrophysical formation channels. 
\end{abstract}



\section{Introduction}

A coalescing binary black hole loses a fraction of its rest mass energy and angular momentum through gravitational radiation. Due to the anisotropic nature of this radiation, the system also loses linear momentum \citep{1961RSPSA.265..109B,PhysRev.128.2471}. Most of this linear momentum is radiated at the instance when two black holes collide, resulting in a recoil or a “kick” of the remnant black hole from its center of mass frame \citep{Campanelli_2007, Gonz_lez_2007,Lousto_2008}. A direct signature of non-linear gravity, the kick is independent of the total mass of black hole binary, and can range from a few tens of $\mathrm{km/s}$ to ${\sim}4000~\mathrm{km/s}$ \citep{PhysRevD.101.024044, mahapatra2023testing}. 

The wide distribution of kick fundamentally impacts the astrophysical formation of black holes across its mass range: from stellar black holes \citep{Gerosa_2019, Mahapatra_2021, mahapatra2024reconstructing,alvarez2024kicking} to intermediate-mass black holes \citep{Holley_Bockelmann_2008} to super-massive black holes \citep{Merritt_2004, 2006MNRAS.372.1540M, Dunn_2020}. All BBH mergers reported so far by LIGO-Virgo-KAGRA Collaborations \citep{AdvLIGORef, AdvVirgoRef, KAGRA, GWTC3} must have produced a remnant black hole that was kicked away from its center of mass by tens to thousands of km/s \citep{Doctor_2021}. The kick of remnant black hole also have novel multi-messenger counterparts for heavy-stellar \citep{McKernan_2019} and super-massive black holes \citep{Bogdanovi_2007}.  

Unlike the mass and the spin of the remnant black hole, a precise measurement of the kick of the remnant black hole has proved to be extremely challenging thus far for generic binaries. There have been inference of the remnant kick based out of constraints on the inspiral parameter. These have been reported for a few LIGO-Virgo-KAGRA Collaboration events \citep{Abbott_2020}. However, except in a case of a few unequal mass binaries \citep{Varma_2022, Mahapatra_2021, bustillo2022gw190412}, the constraints on kick have been much broader than other parameters and require specific binary orientation for further improvement \citep{Calder_n_Bustillo_2018}.

{The goal of our study is to investigate whether multiband observations of binary black holes \citep{2016PhRvL.116w1102S, WP-Multiband, 2020NatAs...4..260J}, i.e. if we measure the  evolution of same source across several orders of frequency spectrum, could inform us about the fate of the remnant black hole.} In this work, we focus primarily on the recovery of the kick vector for the heaviest black holes reported by the LIGO-Virgo-KAGRA Collaboration, such as GW190521 \citep{gw190521}. As we demonstrate, the multiband observations surpass any previous techniques to constrain kick in the current and upcoming gravitational-wave detector network, thus opening new avenues for multi-messenger follow-up and confirming formation channels. In Section \ref{sec:tale}, we describe the setup of our study, and in Section \ref{sec:method} we summarize the tools utilized to conduct the investigation. Section \ref{sec:results} highlights the key results and in Section \ref{sec:discussions} we discuss the astrophysical implications for `lite' intermediate-mass black holes such as GW190521.

\section{Tale of two black holes}
\label{sec:tale}
\begin{figure}[t!]
\        \includegraphics[scale=0.64,trim = {20 0 0 0}]{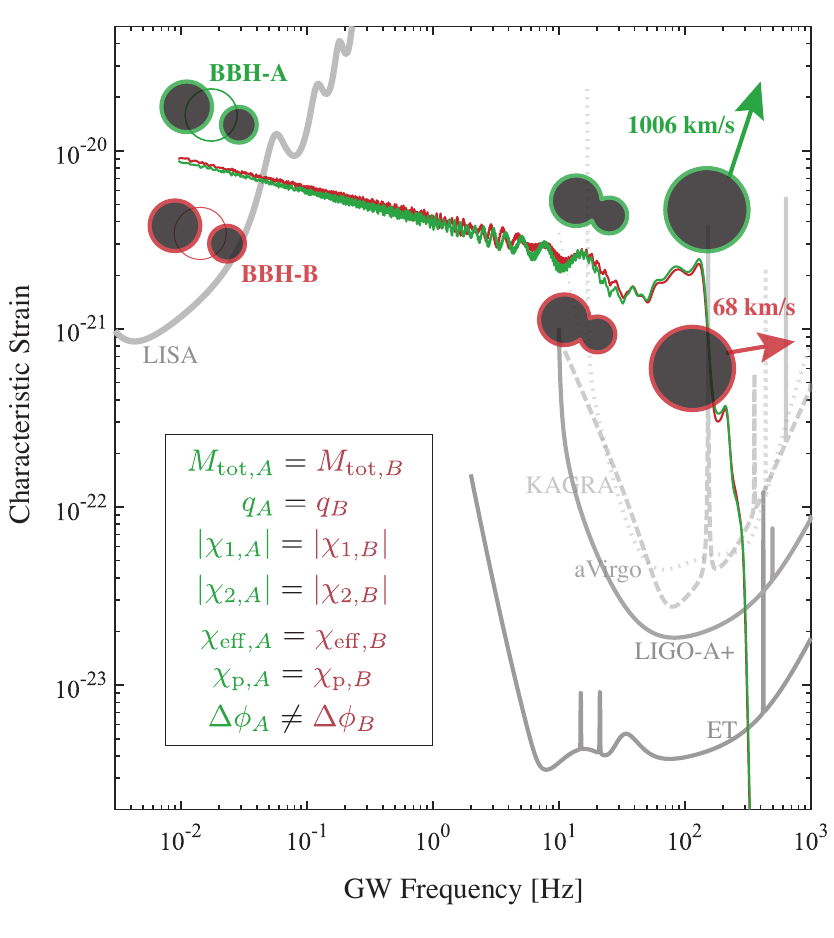}
        \caption{A GW190521-like BBH system, with recoil kick configurations; 68 km/s (red curve) 1006 km/s (green curve). Both systems differ only in their in-plane spin angles.} 
        \label{fig:strain}
    \end{figure}

To conduct our investigation of the kick measurement, we analyze gravitational-wave signals from two distinct binary black hole systems: BBH-A and BBH-B. Figure \ref{fig:strain} shows the multiband frequency evolution of these two binaries.  Both the binaries have source properties similar to that of GW190521 - the heaviest binary black hole merger measured so far, whose remnant provided the first direct measurement of an intermediate-mass black hole \citep{gw190521}. The parameters of the two binaries are shown in Table \ref{tab:injection}. The high spin magnitudes and the spin directions of the two pre-merger black holes are consistent with hierarchical merger scenario of GW190521 \citep{Abbott_2020, Kimball_2021, Gerosa_2021, mahapatra2024predictions,mahapatra2024reconstructing}.

From an astrophysical perspective, both BBH-A and BBH-B are identical. Their primary and secondary black holes have the same exact masses, $m_{1,A}=m_{1,B},~m_{2,A}=m_{2,B}$, and spin magnitudes $|{\chi}_{1,A}|=|{\chi}_{1,B}|,~|{\chi}_{2,A}|=|{\chi}_{2,B}|$. Furthermore, both the binaries have the same effective inspiral spin parameter \citep{PhysRevD.64.124013}, $\chi_{\mathrm{eff},A} = \chi_{\mathrm{eff},B}$, and the same effective precession spin parameter \citep{chip}, $\chi_{\mathrm{p},A} = \chi_{\mathrm{p},B}$. After the merger, both these binaries produce a final black hole of nearly similar mass $m_{\mathrm{f},A} \simeq m_{\mathrm{f},B}$ and spin $|\chi_{\mathrm{f},A}|\simeq|\chi_{\mathrm{f},B}|$ \citep{Varma_2019}. The post-merger parameters of the two binaries are shown in Table \ref{tab:injection}.

From a gravitational-wave data-analysis perspective, both the binaries will record the same signal-to-noise ratio (SNR) in a given gravitational-wave detector. Therefore, the constraints on their pre-merger parameters, especially chirp mass and the effective inspiral spin, will be similar in a given gravitational-wave detector. Therefore, from an inspiral-merger analysis, these two binaries should be indistinguishable from each other for all practical purpose. In earth-based detectors, both the binaries will imprint similar set of ringdown modes, albeit the relative phase and amplitudes of the individual modes will have difference \citep{PhysRevD.77.044031}. 

However, both the binaries have a very different fate. The final black hole produced by BBH-A is kicked from its center of mass at $|v_{\mathrm{f},A}| = 1006~\mathrm{km/s}$, while BBH-B is kicked at $|v_{\mathrm{f},B}| = 68~\mathrm{km/s}$. Even more interesting is that BBH-A is kicked almost entirely in the direction perpendicular to the orbital plane, while BBH-B is kicked in the direction of the orbital plane. The  kick components for both the binaries are shown in \ref{tab:recovery}.

The dramatic difference in the kick vector of these binaries originates from the azimuthal direction of black hole spins $(\phi_{1}, \phi_{2})$. This subtle effect changes how the  two black hole horizons are oriented with respect to binary separation vector just before their merger. We have fine tuned to determine set of azimuthal angles of black hole spins that result in a high kick (BBH-A) vs. low kick (BBH-B) scenario. Since the spin evolution can be defined from the earliest stages of inspiral, the fate of the remnant black hole is sealed billions of years before merger: one remains in the host environment (BBH-B), other becomes a rogue black hole (BBH-A).   

In this study, we use BBH-A as a proxy for the ``high-kick'' type of binaries, and BBH-B as the ``low-kick'' type of binaries. With both our example binaries being nearly identical, comparing BBH-A (high-kick) with BBH-B (low-kick) represents the most challenging astrophysical scenario for distinguishing kick. 
As we detail in the section below, we recover the remnant kick vector for both these binaries at different timescales of their evolution and for various combinations of multiband detector networks.

\begin{table}[t!]
\caption{\label{tab:injection}
Pre-merger and post-merger parameters of two binary systems investigated in this study. BBH-A refers to high kick and BBH-B refers to low kick. The spin angles are defined at a reference frequency of 5 Hz.}
\begin{ruledtabular}
\begin{tabular}{lll}
{\bf Parameter} & {\bf BBH-A} & {\bf BBH-B} \\
\hline
Primary mass, $m_1~(M_\odot)$ & 90 & 90  \\
Primary spin mag., $|\chi_1|$ & 0.7 & 0.7  \\
Primary spin tilt angle, $\theta_1$ & $23.5^{\circ}$ & $23.5^{\circ}$ \\
Primary spin azimuthal angle, $\phi_1$ & $189.0^{\circ}$ & $149.5^{\circ}$ \\
\hline
Secondary mass, $m_2~(M_\odot)$ & 60 & 60  \\
Secondary spin mag., $|\chi_2|$ & 0.7 & 0.7  \\
Secondary spin tilt angle, $\theta_2$ &$100.8^{\circ}$ & $100.8^{\circ}$ \\
Secondary spin azimuthal angle, $\phi_2$ &$288.7^{\circ}$ & $159.8^{\circ}$ \\
\hline
Mass-ratio, $m_2/m_1 \leq 1$ & 0.67 & 0.67 \\
Effective inspiral spin, $\chi_\mathrm{eff}$ & 0.33 & 0.33  \\
Effective precession spin, $\chi_\mathrm{p}$ & 0.43 & 0.43  \\
\hline
Final mass,$m_\mathrm{f}~(M_\odot)$ & {141.31} & {141.16}  \\
Final spin mag., $|\chi_\mathrm{f}|$ & {0.82} & {0.83}  \\
Final kick mag., $|v_\mathrm{f}|$~(km/s)  & 1006.10 & 68.65 \\
\end{tabular}
\end{ruledtabular}
\end{table}

\section{Multiband Inference of Kick}  
\label{sec:method}
With masses similar to GW190521, both BBH-A and BBH-B offer strong prospects for multiband gravitational-wave observations between the upcoming ESA/NASA space mission LISA and earth-based detectors \citep{2020NatAs...4..260J}. The early inspiral (years before the merger) will be measured by LISA around 10s of milli-Hz, while the late-inspiral, merger and ringdown (less than a second before the merger) will be measured around 10s of Hz by LIGO-like detectors. 

The remnant kick strongly depends on the individual spins of the two pre-merger black holes. In this context, the long signal duration in LISA is particularly effective in probing the evolution of black hole spin precession cycles at post-Newtonian regime \citep{spinprecess}. On the other hand, the gravitational burst measured in LIGO-like detectors can infer the higher harmonics radiated due to black hole spins in the non-linear regime \citep{Pekowsky:2012sr}. Additionally, LISA will provide much stronger constraints on the chirp mass of binaries in such IMBH mass-range \citep{1994PhRvD..49.2658C}.

As shown in the top-left side of Figure \ref{fig:strain}, we start by analyzing BBH-A and BBH-B around 5 years before their merger. At this stage, the binaries have an orbital period of ${\sim}1000$ seconds, emitting gravitational waves with monotonically increasing frequency in the range $0.01-0.1~\mathrm{Hz}$. When these binaries leave the LISA band, they will be about an hour away from the merger. At a luminosity distance of 1~Gpc, the signal-to-noise ratio (SNR) for these binaries in the LISA band will be ${\approx}8$.

Just about a second before the merger, both BBH-A and BBH-B will be visible to various earth-based detectors within their frequency range spanning ${\sim}5-200~\mathrm{Hz}$. 
In our analysis, we consider two scenarios for earth-based detectors in the 2030s. The first scenario is the existing facilities of LIGO Hanford, LIGO Livingston \citep{AdvLIGO}, VIRGO \citep{AdvVirgo}, LIGO India \citep{UNNIKRISHNAN_2013} and KAGRA \citep{KAGRA}. We refer this network collectively as Earth-2G. In this scenario, we utilize LIGO noise at \texttt{A+} design. At 1 Gpc, the SNR for these binaries in Earth-2G band will be ${\approx}80$. The second scenario is that of a third-generation detector such as Einstein Telescope (ET) \citep{ETPunturo_2010}. We refer this network as Earth-3G. At 1 Gpc, the SNR for these binaries in Earth-3G band will be ${\approx} 800$. We expect a similar SNR and recovery for the injections in Cosmic Explorer \citep{reitze2019cosmic}.

\begin{figure*}[t!]
    \centering
  \includegraphics[scale=0.82,trim = {10 10 0 0}]{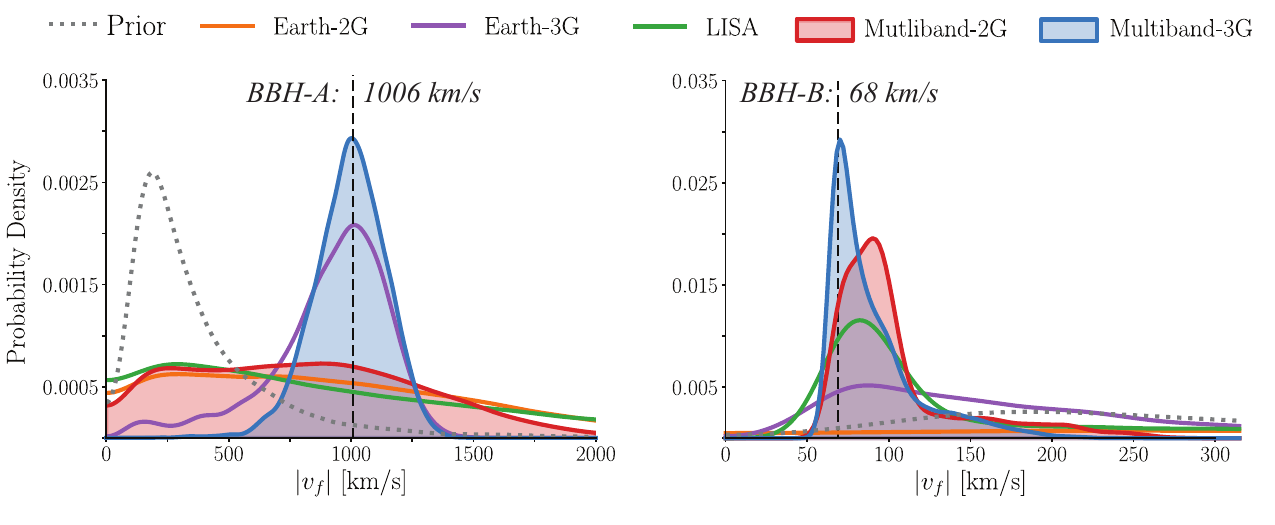}
    \caption{Posteriors on the recovery of kick magnitude for BBH-A (1006 km/s, left plot) and BBH-B (68 km/s, right plot). Shaded plots with thick lines refer to refer to recovery from  multiband networks. Thick lines refer to single band networks. The prior is shown in grey dotted line.  }
    \label{fig:fig2}
\end{figure*}

For injection and recovery of simulated signals in both LISA and earth-based detectors, we use the phenomenological frequency-domain model \verb=IMRPhenomXPHM=   \citep{Pratten_2021}. This waveform model incorporates gravitational-wave signals emitted by quasi-circular precessing binary black holes and includes multipoles beyond the dominant quadruple in the precessing frame. From available waveforms models, we chose this because of the avaibility of higher order modes, including that from spin precession, across the frequency evolution from the LISA band (mHz) to earth-based detectors (kHz). The black hole spin vectors are defined at a reference frequency of $5\mathrm{Hz}$. No spin-weighted spherical harmonic is excluded in either the injection or the inference. The black hole masses are first measured in the detector frame and then converted to their source frame assuming Planck cosmology \citep{2018arXiv180706209P}. We put both the binaries at luminosity distance of 1 Gpc, sky-location with right accession = $126^\circ$ and declination = $-71^\circ$, and inclination from the line of sight set to $30^\circ$.

To calculate the recoil kick vector, we use \verb=surfinBH= \citep{Varma_2019} on our recovered posteriors. \verb|surfinBH| is a module to use a fit trained on numerical relativity simulations to infer remnant properties of binary black holes. It accepts mass ratio, total mass, and individual spin vectors of the progenitor black holes in the binary at a reference frequency, and outputs the properties of the remnant black hoke such as the post-merger mass, the post-merger spin vector and the recoil kick vector. From the available phenomenological fits, we use this model because its kick is calibrated directly the ringdown modes extrapolated in numerical relativity simulations. While our pre-merger and post-merger models  belong to different family of approximants, they represent the best choices in their respective frequency regimes. A systematic comparison of kick estimates for different family of models has been conducted by \cite{Borchers_2023}. 

The injection analysis is carried out using PyCBC \citep{PyCBCInf}, an open source python-based toolkit implementing Bayesian Inference for gravitational wave signals. For studying the signal in the two earth-based scenarios, we run a Bayesian inference package \texttt{PyCBC-Inf}. In the absence of any information from LISA for terrestrial detectors, we use uniform priors across the 14D intrinsic and extrinsic parameters of BBH. This includes the mass constraints between $50M_\odot$ to $1000M_{\odot}$, mass-ratio between 1 and 10, and uniform distribution of the black hole spin magnitude and directions. Such a uniform distribution of mass ratio and spins results in the prior black hole kicks between 0 to 5000 km/s, however, it is not uniformly placed.  

In the case of multiband observations, we replace the priors for the Bayesian inference in earth-based detectors with the Gaussian variances obtained from the Fisher Matrix analysis in LISA. To compute variances in LISA, we utilize the \texttt{Gwbench} formalism \citep{Borhanian_2021}. As shown by several past studies \citep{2016PhRvL.117e1102V, WP-Multiband, Randall_2021}, such an approach of combining LISA's early constraints can significantly improve the parameter estimation in earth-based detectors. More recent work have considered a fully Bayesian inference for the multiband signals in LISA \citep{Buscicchio_2021}.

We consider two multiband scenarios: (i) the conservative scenario of only Earth-2G detectors operating during LISA  mission lifetime, and (ii) the optimistic scenario of Earth-3G detector operating at design sensitivity during LISA's mission duration. We report the measurement on the black hole kicks of BBH-A and BBH-B for five single band and multiband network combinations: 
\begin{enumerate}
    \item LISA: 5-year mission lifetime 
    \item Earth-2G: Hanford - Livingston - Virgo - India - KAGRA (HLVIK), with HLI at A+ sensitivity 
    \item Earth-3G: Einstein Telescope at design sensitivity
    \item Multiband-2G: LISA and Earth-2G network
    \item Multiband-3G: LISA and Earth-3G network
\end{enumerate}

\begin{table*}[t!]
    \caption{\label{tab:recovery}
        Injection and recovery of the kick magnitude and kick vector projections for BBH-A and BBH-B for the five detector combinations. Reported are median Values with 90\% credible intervals that include statistical errors. The columns show the recovery of the final kick magnitude $|v_f|$, the component of final kick perpendicular to the orbital plane $v_{\parallel}$, the component of final kick perpendicular to plane $v_{\perp}$, and the in plane kick components $v_x$ and $v_y$ respectively. The rows show the injected values, as well as the values recovered for each detector combination.}
        \begin{ruledtabular}
        \begin{tabular}{lccccc}
        &{\bf  Magnitude}&{\bf In-plane } &{\bf Perpendicular }&{\bf Azimuthal  $x$} &{\bf Azimuthal  $y$} \\
        
          &{ $|v_\mathrm{f}|$ [km/s]}&{ $v_\mathrm{\parallel}$ [km/s]}&{$v_\mathrm{\perp}$ [km/s]} &{$v_x$ [km/s]}&{$v_y$ [km/s]} \\
        \hline
        \textbf{Injection Values} & & & & & \\
        \ \ BBH-A & 1006.10 & 203.56 & 985.3 & -121.0 & -163.69 \\
        \ \ BBH-B & 68.65 & 67.8 & 10.5 & 67.49 & -6.94 \\
        \hline

        \textbf{Earth-2G} & & & & & \\
        \vspace{0.03in}
        
        \ \ BBH-A & $856_{-662}^{+948}$ & $108_{-76}^{+199}$ & $-8_{-1832}^{+1760}$ & $-4_{-165}^{+182}$ & $6_{-186}^{+182}$ \\
        \ \ BBH-B & $617_{-481}^{+523}$ & $104_{-74}^{+178}$ &  $-24_{-1567}^{+1652}$ & $-1_{-160}^{+166}$ & $4_{-182}^{+155}$ \\

        \hline
        
        \textbf{Earth-3G} & & & & & \\
        \vspace{0.03in}
        
        \ \ BBH-A & $958_{-377}^{+209}$ & $197_{-84}^{+25}$ & $937_{-572}^{+261}$ & $-138_{-40}^{+58}$ & $-132_{-41}^{+97}$\\
        \ \ BBH-B & $149_{-82}^{+176}$ & $73_{-40}^{+41}$ & $10_{-217}^{+360}$ & $39_{-97}^{+61}$ & $1_{-60}^{+94}$ \\

        \hline

        \textbf{LISA} & & & & & \\
         \vspace{0.03in}
       \ \ BBH-A & $764_{-589}^{+1134}$ & $147_{-108}^{+348}$ & $-22_{-1853}^{+1852}$ & $7_{-292}^{+315}$ & $-3_{-256}^{+268}$ \\
        \ \ BBH-B & $95_{-26}^{+212}$ & $68_{-34}^{+36}$ & $20_{-89}^{+330}$ & $44_{-99}^{+34}$ & $-6_{-62}^{+89}$ \\

        \hline

        \textbf{Multiband-2G} & & & & & \\
        \vspace{0.03in}
        \ \ BBH-A & $794_{-591}^{+662}$ & $150_{-107}^{+139}$ & $334_{-1617}^{+1175}$ & $-35_{-156}^{+187}$ & $-47_{-183}^{+226}$ \\
        \ \ BBH-B & $92_{-22}^{+74}$ & $69_{-36}^{+14}$ & $1_{-66}^{+195}$ & $65_{-62}^{+13}$ & $-9_{-25}^{+47}$ \\

        \hline

        \textbf{Multiband-3G} & & & & & \\
        \vspace{0.03in}
        \ \ {BBH-A} & $1001_{-191}^{+174}$ & $200_{-44}^{+18}$ & $981_{-249}^{+220}$ & $-132_{-37}^{+47}$ & $-148_{-27}^{+56}$ \\
        \ \ {BBH-B} & $80_{-14}^{+51}$ & $62_{-29}^{+20}$ & $23_{-75}^{+127}$  & $62_{-44}^{+18}$ & $-3_{-18}^{+29}$ \\

        \end{tabular}
\end{ruledtabular}
\end{table*}

To summarize our multiband method of inferring kick: we start by analyzing BBH-A and BBH-B a few years before the merger, when they emit gravitational waves in the LISA band. Using Fisher Matrix analysis, we obtain posteriors on the intrinsic (masses, spins) and extrinsic parameters (distance, sky-location, inclination) of these binaries. Next, when the signal is visible in the earth-based detectors (both conservative and optimistic scenarios), we run the parameter estimation using \texttt{PyCBC-Inf} where the priors are informed by the posteriors we obtained from LISA. The final step in the workflow is to employ the numerical relativity fit \verb=surfinBH= to infer the recoil kick vector from the constraints on binary black hole inspiral configuration.

\begin{figure*}[t!]
    \centering
  \includegraphics[scale=0.91,trim = {20 0 0 0}]{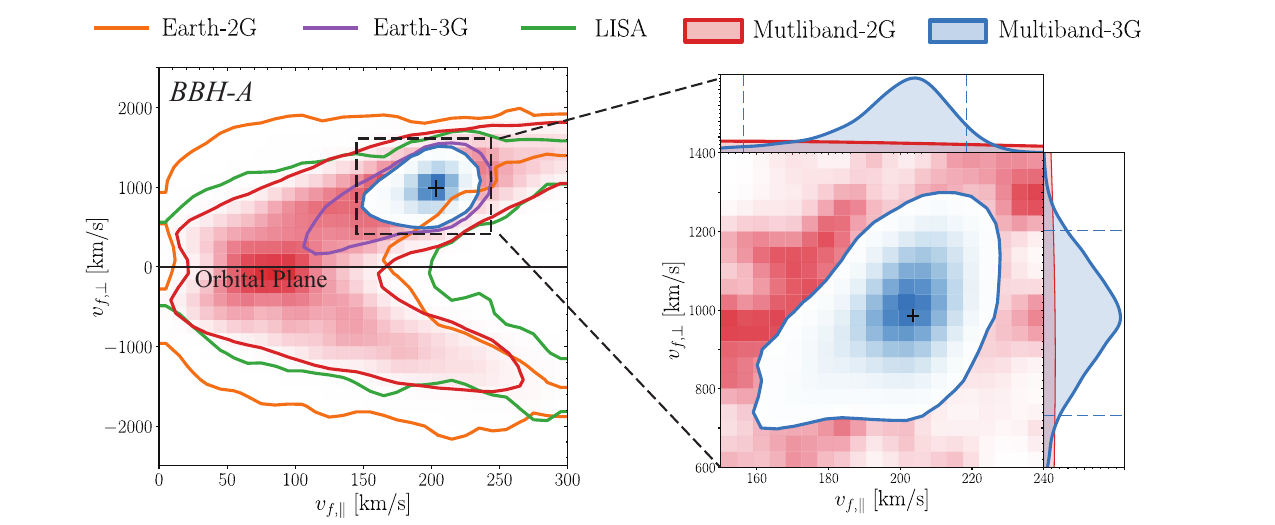}
  \includegraphics[scale=0.91,trim = {20 0 0 20}]{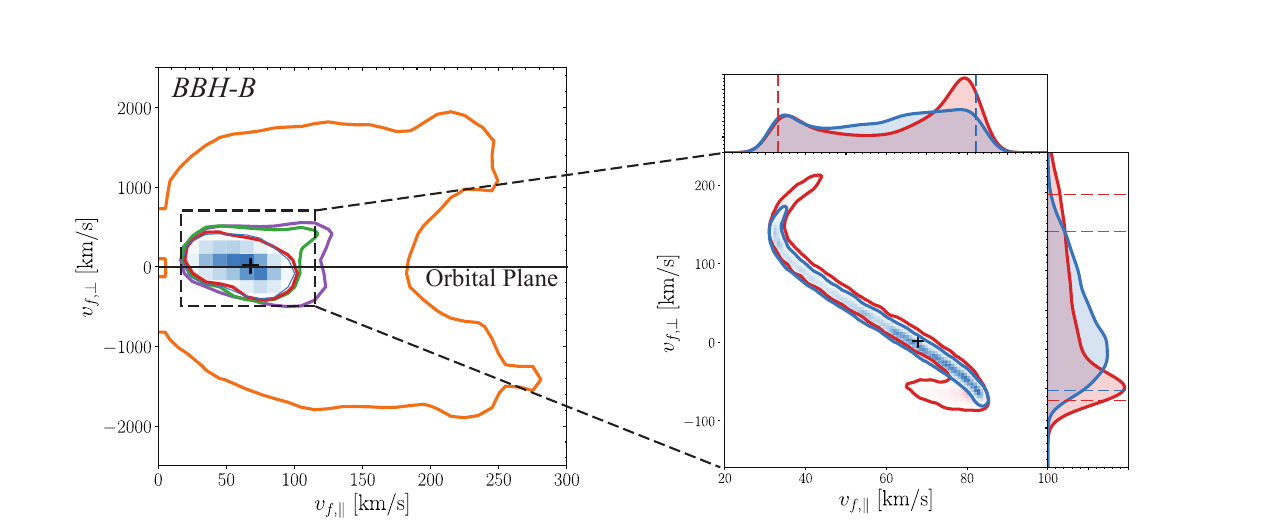}
    \caption{
    Recovery of the projection of kick vector along the orbital plane (xy-plane) and perpendicular to the orbital plane (z-axis) for BBH-A (1006~km/s, \textit{Top}) and BBH-B  (68~km/s, \textit{Bottom}). The contour lines refer to the 90\% confidence interval. The zoomed-in plots on the right shows the recovery around the injection with multiband recoveries. 
    The black horizontal line shows the orbital plane in the "edge-on" view, and the injection is shown with `\texttt{+}' for both the cases.}
    \label{fig:Fig3}
\end{figure*}

\section{Results}
\label{sec:results}

\begin{figure*}[t!]
    \centering
  \includegraphics[scale=0.91,trim = {20 0 0 0}]{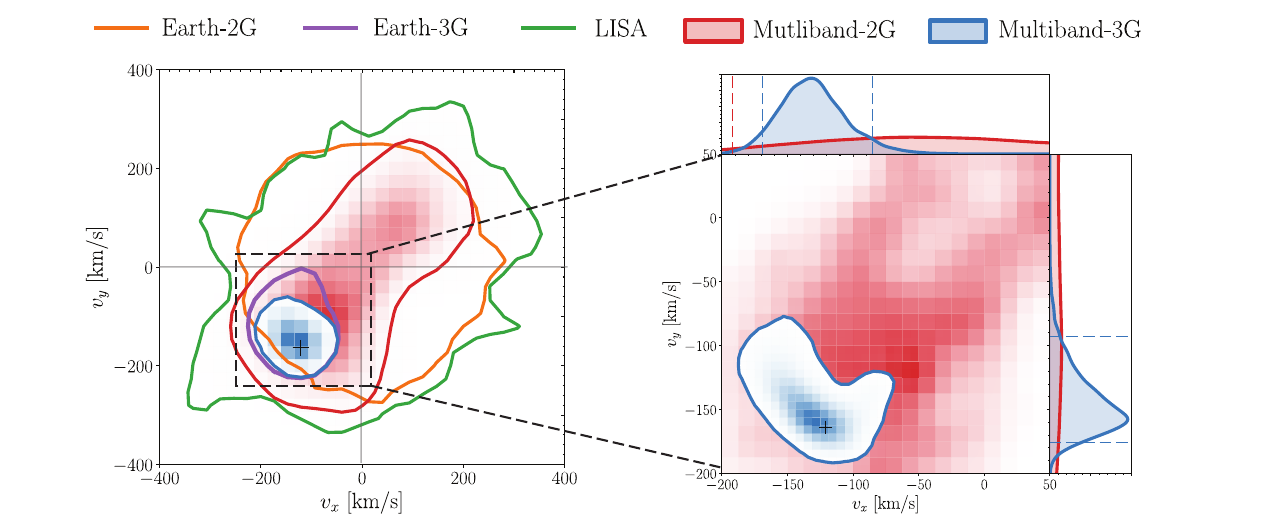}
  \includegraphics[scale=0.91,trim = {20 0 0 10}]{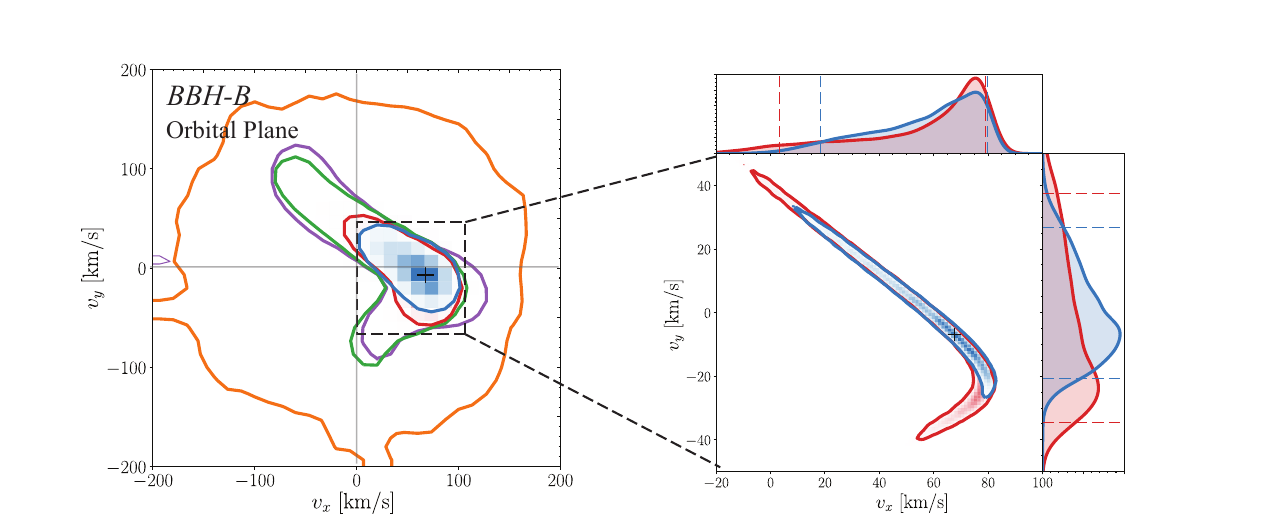}
    \caption{Recovery of the projection of kick vector in the orbital plane (xy-plane) for BBH-A (1006~km/s, \textit{Top}) and BBH-B (68~km/s, \textit{Bottom}). The contour lines refer to the $90\%$ confidence interval. The zoomed-in plots on the right show the recovery around the injection with multiband recoveries. The black horizontal lines are the $x=0$ and $y=0$ lines in the orbital plane in the ``face-on" view, and the injection is shown with `\texttt{+}' for both cases.}
    \label{fig:Fig4}
\end{figure*}

The overall results on the recovery for kick for BBH-A and BBH-BH across multiband networks are summarized in Table \ref{tab:recovery}. The table reports the 90\% confidence interval in posterior distribution for kick magnitude $(|v_\mathrm{f}|)$, the polar angle of kick ($\theta_v$, where $\theta_v=0$ is pointing towards orbital angular momentum of the binary) and the azimuthal direction of the kick in the orbital plane ($\phi_v$, where $\phi_v=0$ is pointing towards the X-axis defined in our coordinate systems for the spin vectors). 

\subsection{Kick Magnitude}
In Figure \ref{fig:fig2}, we report the posteriors of the kick magnitude for BBH-A (high-kick magnitude: 1006~km/s) and BBH-B (low-kick magnitude: 68~km/s) recovered for our five detector combinations. Note the prior on the kick magnitude for a uniform distribution of binary black hole spin vectors across moderate mass-ratios $(m_1/m_2 < 10)$ spans up to ${\sim}5000$~km/s, with a peak at around 100~km/s. 

In the current generation of gravitational-wave facilities (Earth-2G), we find that the recovery of kick magnitude is not skewed by the underlying distribution of prior. However, as shown in Figure \ref{fig:fig2}, Earth-2G network is not sensitive enough to distinguish between a high-kick case (BBH-A) and a low-kick case (BBH-B). For both these cases, we find that the 90\% confidence interval spans from a few 100~km/s to around 1800~km/s. The magnitude of the low-kick case (BBH-B) is not recovered even with 90\% bounds. Therefore, we deem the current set of gravitational-wave observatories not ideal for reliable kick magnitude measurements for generic binaries.   

For the next-generation detector ET at its design sensitivity, there is a notable improvement in the measurement of kick magnitude. In particular, ET can distinguish between a high-kick case (BBH-A) and a low-kick case (BBH-B). However, the posterior distribution is fairly broad, especially for BBH-B. Within 90\% confidence intervals, the BBH-B can be up to 350~km/s - almost a factor or 6 more than the injected value of kick.

A standalone observation of inspiral in LISA provides a better measurement of the kick magnitude than the HLVIK network on Earth. As shown in Table \ref{tab:recovery}, LISA can well distinguish between a high-kick case (BBH-A) and a low-kick case (BBH-B) with a 90\% confidence interval. But the posterior samples, as seen in Figure \ref{fig:fig2}, are fairly broad for the BBH-A case. This is still promising, as LISA has no measurement of the late inspiral and merger and yet can predict the post-merger fate of a binary. 

A multiband observation between LISA and HLVIK (Multiband-2G, conservative case) is particularly promising. Adding on the LISA's recovery of kick magnitude, the HLVIK constraints with priors informed by LISA reduces the error bars for the low-kick case of BBH-B substantially, under few tens of km/s. However, the error bars on the high-kick case of BBH-A still remain fairly high.

The best improvement comes from multiband observation between LISA and ET (Multiband-3G, optimistic case).  The 90\% confidence interval for kick posterior BBH-A is within 100~km/s error bards and BBH-B is in tens of km/s. The effect of LISA priors is particular notable for the low-kick case BBH-B, where compared to Earth-3G, the kick magnitude has now reduced by a factor of ${\sim}4$x.

\subsection{Kick Direction}

We constrain the kick direction in our study in two ways - firstly we constrain the direction of kick while looking edge-on to the orbital plane and recovering the component of the kick vector perpendicular and along the plane (Figure \ref{fig:Fig3}), and secondly, we look at the constraint in kick vector in the orbital plan ``face-on" (Figure \ref{fig:Fig4}). The numerical values of the median of the posterior for all these components of kick vector with their $90\%$ credible intervals are recorded in Table \ref{tab:recovery}.

Figure \ref{fig:Fig3} shows the $90\%$ credible interval contours of the recovery of the kick vector components along the orbital plane (perpendicular to plane) along the horizontal axis (vertical axis) for BBH-A and BBH-B in the top and bottom panel respectively. The zoomed-in, filled in plot in the right column for both BBH-A and BBH-B shows the multiband case (multiband-2g and multiband-3g). The injections are indicated in both cases, and we see that BBH-A has a significant component in both directions, majorly along the direction perpendicular to the plane, whereas BBH-B has no component perpendicular to the plane, and the kick is majorly along the plane of the orbit. These two contrasting cases of the direction of the kick vector with respect to the orbital plane makes this an important check for recovery of the kick direction with all five of our detector combinations and comparing if our best detector combinations perform consistently across these two cases.

The Earth-2G detector configuration, on its own, cannot differentiate between the two cases and has a symmetric distribution about the orbital plane for both cases. This is also seen in Table \ref{tab:recovery}, in columns labelled $v_{\parallel}$ and $v_{\perp}$ where the credible region is very broad. Earth-3G manages to place a tight constraint for both BBH-A and BBH-B, doing significantly better as a standalone detector configuration than Earth-2G, as is seen from Figure \ref{fig:Fig3}. 

LISA manages to constrain the kick direction better for BBH-B - managing to clearly distinguish between the zero perpendicular-to-plane kick case with a case which has a component in both the directions. It is interesting to note that LISA manages to constraint this for BBH-B marginally better than even Earth-3G. Multibanding Earth-2G with LISA result in marginal improvement for BBH-A since both these detectors have equally broad, bimodal posteriors for BBH-A, and so their combination does only marginally better. Multibanding Earth-3G with LISA gives us the tightest constraint on the kick direction out of all five detector combinations we have considered for both BBH-A and BBH-B. This can also be seen in the zoomed in plots in the right column, that multiband-3G manages to recover the injection well within the $90\%$ credible region for both cases with an accuracy of few tens of km/s accuracy especially in the in-plane component.

Figure \ref{fig:Fig4} shows the constraint on the $90\%$ credible region of the kick vectors in the orbital-plane, in the x and y direction indicated by the respective final kick in that direction. The top row shows the case for BBH-A, with the left panel showing all five detector combinations and the right panel showing the enlarged plot specifically for the multiband case. Similarly, the bottom row shows recovery for BBH-B. The injections are indicated in both cases. 
We see that while BBH-A has an almost equal component of the final kick in both x and y directions, BBH-B has a near zero value for the y-component of the final kick vector. Similar to Figure \ref{fig:Fig3}, this presents two different cases in terms of the final kick vector direction.

Although Earth-2G recovers the injection within the posterior's $90\%$ credible interval, the posterior spans a broad area for both the cases, as can also be seen from Table \ref{tab:recovery}, in columns labelled $v_{\parallel}$ and $v_{\perp}$ as well as in Figure \ref{fig:Fig4}. Earth-3G performs marginally better than Earth-2G for BBH-A and significantly better for BBH-B. LISA performs worse than Earth-2G for BBH-A, but does significantly better for BBH-B - where it manages to perform better than Earth-3G.
Multibanding earth-2G with LISA (multiband-2G) performs marginally better than earth-3g alone, where the combination of earth-2g with LISA manages to break enough degeneracies to perform better than a third generation detector working alone. For BBH-B, multiband-2g does a better job than LISA alone in constraining the in plane components. The tighter constraint for BBH-B using LISA alone results in an improved constraint for multiband-2g. Multiband-3G gives us our best constraint on the in plane kick components of the final kick, manging to recover the injection for both BBH-A and BBH-B, well within the credible interval with an accuracy of about a few tens of km/s for BBH-B and a about a hundred km/s for BBH-A. This is evident in Figure \ref{fig:Fig4} as well as the last two rows of Table \ref{tab:recovery}.

\section{Astrophysical Implications}
\label{sec:discussions}

The astrophysical formation channel for a binary with primary and secondary black holes in the pair-instability supernovae mass-gap $({\sim}60~M_\odot$ to ${\sim} 120~ M_\odot)$ is widely debated. Several such confirmed and candidate black holes have been detected over the past few years with gravitational-wave data of the LIGO and Virgo detectors \citep{2020ApJ...900...80U, gw190521, O3IMBH, GWTC3}. A list of prominent formation channels that can produce black hole in the mass-gap are reviewed in \cite{Abbott_2020}. The setup to populate black holes in the mass-gap can be broadly divided into two categories: requires multiple generation of mergers of lower-mass black holes (globular clusters), with possible mass accretion and gas dynamics (AGN disks), vs. directly collapses to a black hole (stellar mergers). The difference between the two categories, and its sub-categories, sets a fundamental imprint on the the distribution on the spin vector of the pre-merger black holes, thus directly impacting the kick vector of the remnant black hole. The former has a possibility of producing high kicks as reported for BBH-A, while latter will more likely produce lower kicks as reported for BBH-B case. 

As we report in Section \ref{sec:results}, the current generation of gravitational-wave detectors at their peak performance sensitivity (Earth-2G) will not be successful in differentiating between the two varieties in the kick distribution, unless the binary is highly unequal. By observing these binaries in the LISA band for a couple of years, we find that it would be sufficient to distinguish between the two categories. In the scenario of joint observations, the inclusion of LISA constraints with the earth-based detectors (Multiband-2G, 3G) will improve the recovery of the kick magnitude to such an extent that we can directly rule out the host environment of the remnant black hole. For example, elliptical galaxies and young star clusters can have escape velocities differing by only a few hundred km/s. This will be critical information to know if the remnant black hole has gone rogue or stays back for a possible another merger or interaction with its environment. 

The need to measure the direction of the kick has become pressing for electromagnetic follow-ups of black holes in the mass-gap. A few weeks after the detection of GW190521, a UV flare was reported in an AGN that was located within the uncertainty of the spatial volume of this event \citep{2020ZTF}. Remnant black holes may produce flares through their interaction with the AGN disk after being kicked away after merger \citep{2024ApJ...966...21T, Kimura_2021}. The flare reported after GW190521 was hypothesized to have been produced by the remnant black hole interacting with the AGN disk after merger \citep{McKernan_2019}. As we describe in Section \ref{sec:tale}, the polar direction of the kick vector can vary widely between BBH-A (perpendicular to the orbital plane) compared to BBH-B (parallel to the orbital plane). While both the cases have kick magnitude less than the escape velocity from AGN disks, if the orbital plane is aligned with the disk, then BBH-A would have lesser likelihood of sustained gas accretion compared to BBH-B. By incorporating the azimuthal direction of the kick, one can create a stronger likelihood to determine if the remnant black hole will interact with the AGN disk.  

As we report in Section \ref{sec:results}, measuring kick direction for more precise AGN follow up is futile with Earth-2G detectors. However, observing with LISA and subsequently with the multiband-2G network provides considerable improvement in limiting the polar direction of the kick within a bimodality that points in either perpendicular directions to the orbital plane (BBH-A). With the optimistic scenario of a joint measurement between LISA and ET (Multiband-3G), the degeneracy is broken and one can exactly know which perpendicular direction the black hole was kicked with respect to the line sight. Furthermore, Multiband-3G has remarkable accuracy in measuring the azimuthal direction for both BBH-A and BBH-B cases. Therefore, it can precisely determine if the flare was more likely for any GW190521 like binary source. 

While GW190521 offers a powerful motivation for measuring kick vector, its astrophysical mergers rates are very rare, almost 200x lower than stellar black hole mergers \citep{Abbott_2022}. Therefore, we expect only a handful of extremely fortunate multiband detections during LISA entire mission lifetime. The multiband rates for such `lite' intermediate-mass black holes are much higher for mid-band (sub-Hertz) detectors on the Moon and in space \citep{Sedda_2020, Jani_2021, Harms_2021, Izumi_2021}.   
Future work will include different astrophysical scenarios for GW190521 and stellar binary systems, including lower spin magnitudes, varying mass-ratios and mid-band detectors. 

\section*{Acknowledgments}
We thank Imre Bartos and Juan Calderon Bustillo for important feedback. KHB acknowledges and appreciates support from NSF NRT-2125764. AHN acknowledges support from NSF grant PHY-2309240. C.C. acknowledges support from NSF grant PHY-2309231. C.C.’s work was carried out at the Jet Propulsion Laboratory,  California Institute of Technology, under a contract with NASA. This work used the resources provided by the Vanderbilt Advanced Computing Center for Research and Education (ACCRE), a collaboratory operated by and for Vanderbilt faculty at Vanderbilt University.

\bibliography{references.bib}

\providecommand{\noopsort}[1]{}\providecommand{\singleletter}[1]{#1}%
\begin{thebibliography}{}
\expandafter\ifx\csname natexlab\endcsname\relax\def\natexlab#1{#1}\fi
\providecommand{\url}[1]{\href{#1}{#1}}

\bibitem[{Aasi {et~al.}(2015)Aasi, Abbott, Abbott, Abbott, Abernathy, Ackley,
  Adams, Adams, Addesso, \& et~al.}]{AdvLIGO}
Aasi, J., Abbott, B.~P., Abbott, R., {et~al.} 2015, Classical and Quantum
  Gravity, 32, 074001.
\newblock \url{http://dx.doi.org/10.1088/0264-9381/32/7/074001}

\bibitem[{{Abbott} {et~al.}(2023){Abbott}, {Abbott}, {Zucker}, {Zweizig}, {Ligo
  Scientific Collaboration}, \& {Kagra Collaboration}}]{GWTC3}
{Abbott}, R., {Abbott}, T.~D., {Zucker}, M.~E., {et~al.} 2023, Physical Review
  X, 13, 041039

\bibitem[{Abbott {et~al.}(2020{\natexlab{a}})Abbott, Zhou, Zhu, Zimmerman,
  Zlochower, Zucker, \& Zweizig}]{Abbott_2020}
Abbott, R., Zhou, Z., Zhu, X.~J., {et~al.} 2020{\natexlab{a}}, The
  Astrophysical Journal Letters, 900, L13.
\newblock \url{http://dx.doi.org/10.3847/2041-8213/aba493}

\bibitem[{Abbott {et~al.}(2020{\natexlab{b}})Abbott, Abbott, Abraham, Acernese,
  Ackley, Adams, Adhikari, Adya, Affeldt, Agathos, \& et~al.}]{gw190521}
Abbott, R., Abbott, T., Abraham, S., {et~al.} 2020{\natexlab{b}}, Physical
  Review Letters, 125, doi:10.1103/physrevlett.125.101102.
\newblock \url{http://dx.doi.org/10.1103/PhysRevLett.125.101102}

\bibitem[{{Abbott} {et~al.}(2022){Abbott}, {Abbott}, {Acernese}, {Ackley},
  {Adams}, {Adhikari}, {Adhikari}, {Adya}, {Affeldt}, {Agarwal}, {Agathos},
  {Agatsuma}, {Aggarwal}, {Aguiar}, {Aiello}, {Ain}, {Ajith}, {Akutsu},
  {Albanesi}, {Allocca}, {Altin}, {Amato}, {Anand}, {Anand}, {Ananyeva}, \&
  {Anderson}}]{O3IMBH}
{Abbott}, R., {Abbott}, T.~D., {Acernese}, F., {et~al.} 2022, Astronomy \&
  Astrophysics, 659, A84

\bibitem[{Abbott {et~al.}(2022)Abbott, Abbott, Acernese, Ackley, Adams,
  Adhikari, Adhikari, Adya, Affeldt, Agarwal, Agathos, Agatsuma, Aggarwal,
  Aguiar, Aiello, Ain, Ajith, Akutsu, Albanesi, Allocca, Altin, Amato, Anand,
  Anand, Ananyeva, Anderson, Anderson, Ando, Andrade, Andres, Andrić,
  Angelova, Ansoldi, Antelis, Antier, Appert, Arai, Arai, Arai, Araki, Araya,
  Araya, Areeda, Arène, Aritomi, Arnaud, Aronson, Arun, Asada, Asali, Ashton,
  Aso, Assiduo, Aston, Astone, Aubin, Austin, Babak, Badaracco, Bader, Badger,
  Bae, Bae, Baer, Bagnasco, Bai, Baiotti, Baird, Bajpai, Ball, Ballardin,
  Ballmer, Balsamo, Baltus, Banagiri, Bankar, Barayoga, Barbieri, Barish,
  Barker, Barneo, Barone, Barr, Barsotti, Barsuglia, Barta, Bartlett, Barton,
  Bartos, Bassiri, Basti, Bawaj, Bayley, Baylor, Bazzan, Bécsy, Bedakihale,
  Bejger, Belahcene, Benedetto, Beniwal, Bennett, Bentley, BenYaala, Bergamin,
  Berger, Bernuzzi, Berry, Bersanetti, Bertolini, Betzwieser, Beveridge,
  Bhandare, Bhardwaj, Bhattacharjee, Bhaumik, Bilenko, Billingsley, Bini,
  Birney, Birnholtz, Biscans, Bischi, Biscoveanu, Bisht, Biswas, Bitossi,
  Bizouard, Blackburn, Blair, Blair, Blair, Bobba, Bode, Boer, Bogaert,
  Boldrini, Bonavena, Bondu, Bonilla, Bonnand, Booker, Boom, Bork, Boschi,
  Bose, Bose, Bossilkov, Boudart, Bouffanais, Bozzi, Bradaschia, Brady,
  Bramley, Branch, Branchesi, Brau, Breschi, Briant, Briggs, Brillet,
  Brinkmann, Brockill, Brooks, Brooks, Brown, Brunett, Bruno, Bruntz, Bryant,
  Bulik, Bulten, Buonanno, Buscicchio, Buskulic, Buy, Byer, Cadonati, Cagnoli,
  Cahillane, Calderón~Bustillo, Callaghan, Callister, Calloni, Cameron, Camp,
  Canepa, Canevarolo, Cannavacciuolo, Cannon, Cao, Cao, Capocasa, Capote,
  Carapella, Carbognani, Carlin, Carney, Carpinelli, Carrillo, Carullo, Carver,
  Casanueva~Diaz, Casentini, Castaldi, Caudill, Cavaglià, Cavalier, Cavalieri,
  Ceasar, Cella, Cerdá-Durán, Cesarini, Chaibi, Chakravarti,
  Chalathadka~Subrahmanya, Champion, Chan, Chan, Chan, Chan, Chan, Chandra,
  Chanial, Chao, Charlton, Chase, Chassande-Mottin, Chatterjee, Chatterjee,
  Chatterjee, Chaturvedi, Chaty, Chatziioannou, Chen, Chen, Chen, Chen, Chen,
  Chen, Chen, Chen, Cheng, Cheong, Cheung, Chia, Chiadini, Chiang, Chiarini,
  Chierici, Chincarini, Chiofalo, Chiummo, Cho, Cho, Choudhary, Choudhary,
  Christensen, Chu, Chu, Chu, Chua, Chung, Ciani, Ciecielag, Cieślar, Cifaldi,
  Ciobanu, Ciolfi, Cipriano, Cirone, Clara, Clark, Clark, Clarke, Clearwater,
  Clesse, Cleva, Coccia, Codazzo, Cohadon, Cohen, Cohen, Colleoni, Collette,
  Colombo, Colpi, Compton, Constancio, Conti, Cooper, Corban, Corbitt,
  Cordero-Carrión, Corezzi, Corley, Cornish, Corre, Corsi, Cortese, Costa,
  Cotesta, Coughlin, Coulon, Countryman, Cousins, Couvares, Coward, Cowart,
  Coyne, Coyne, Creighton, Creighton, Criswell, Croquette, Crowder, Cudell,
  Cullen, Cumming, Cummings, Cunningham, Cuoco, Curyło, Dabadie, Dal~Canton,
  Dall’Osso, Dálya, Dana, Daneshgaran~Bajastani, D’Angelo, Danilishin,
  D’Antonio, Danzmann, Darsow-Fromm, Dasgupta, Datrier, Datta, Dattilo, Dave,
  Davier, Davies, Davis, Davis, Daw, Dean, DeBra, Deenadayalan, Degallaix,
  De~Laurentis, Deléglise, Del~Favero, De~Lillo, De~Lillo, Del~Pozzo,
  De~Marchi, De~Matteis, D’Emilio, Demos, Dent, Depasse, De~Pietri, De~Rosa,
  De~Rossi, De~Salvo, De~Simone, Dhurandhar, Díaz, Diaz-Ortiz, Didio,
  Dietrich, Di~Fiore, Di~Fronzo, Di~Giorgio, Di~Giovanni, Di~Giovanni,
  Di~Girolamo, Di~Lieto, Ding, Di~Pace, Di~Palma, Di~Renzo, Divakarla,
  Dmitriev, Doctor, D’Onofrio, Donovan, Dooley, Doravari, Dorrington, Drago,
  Driggers, Drori, Ducoin, Dupej, Durante, D’Urso, Duverne, Dwyer, Eassa,
  Easter, Ebersold, Eckhardt, Eddolls, Edelman, Edo, Edy, Effler, Eguchi,
  Eichholz, Eikenberry, Eisenmann, Eisenstein, Ejlli, Engelby, Enomoto, Errico,
  Essick, Estellés, Estevez, Etienne, Etzel, Evans, Evans, Ewing, Fafone,
  Fair, Fairhurst, Farah, Farinon, Farr, Farr, Farrow, Fauchon-Jones, Favaro,
  Favata, Fays, Fazio, Feicht, Fejer, Fenyvesi, Ferguson, Fernandez-Galiana,
  Ferrante, Ferreira, Fidecaro, Figura, Fiori, Fishbach, Fisher, Fittipaldi,
  Fiumara, Flaminio, Floden, Fong, Font, Fornal, Forsyth, Franke, Frasca,
  Frasconi, Frederick, Freed, Frei, Freise, Frey, Fritschel, Frolov, Fronzé,
  Fujii, Fujikawa, Fukunaga, Fukushima, Fulda, Fyffe, Gabbard, Gadre, Gair,
  Gais, Galaudage, Gamba, Ganapathy, Ganguly, Gao, Gaonkar, Garaventa,
  García-Núñez, García-Quirós, Garufi, Gateley, Gaudio, Gayathri, Ge,
  Gemme, Gennai, George, Gerberding, Gergely, Gewecke, Ghonge, Ghosh, Ghosh,
  Ghosh, Ghosh, Giacomazzo, Giacoppo, Giaime, Giardina, Gibson, Gier, Giesler,
  Giri, Gissi, Glanzer, Gleckl, Godwin, Goetz, Goetz, Gohlke, Goncharov,
  González, Gopakumar, Gosselin, Gouaty, Gould, Grace, Grado, Granata,
  Granata, Grant, Gras, Grassia, Gray, Gray, Greco, Green, Green, Gretarsson,
  Gretarsson, Griffith, Griffiths, Griggs, Grignani, Grimaldi, Grimm, Grote,
  Grunewald, Gruning, Guerra, Guidi, Guimaraes, Guixé, Gulati, Guo, Guo,
  Gupta, Gupta, Gupta, Gustafson, Gustafson, Guzman, Ha, Haegel, Hagiwara,
  Haino, Halim, Hall, Hamilton, Hammond, Han, Haney, Hanks, Hanna, Hannam,
  Hannuksela, Hansen, Hansen, Hanson, Harder, Hardwick, Haris, Harms, Harry,
  Harry, Hartwig, Hasegawa, Haskell, Hasskew, Haster, Hattori, Haughian,
  Hayakawa, Hayama, Hayes, Healy, Heidmann, Heidt, Heintze, Heinze, Heinzel,
  Heitmann, Hellman, Hello, Helmling-Cornell, Hemming, Hendry, Heng, Hennes,
  Hennig, Hennig, Hernandez, Hernandez~Vivanco, Heurs, Hild, Hill, Himemoto,
  Hines, Hiranuma, Hirata, Hirose, Hochheim, Hofman, Hohmann, Holcomb, Holland,
  Hollows, Holmes, Holt, Holz, Hong, Hopkins, Hough, Hourihane, Howell, Hoy,
  Hoyland, Hreibi, Hsieh, Hsu, Huang, Huang, Huang, Huang, Huang, Huang,
  Hübner, Huddart, Hughey, Hui, Hui, Husa, Huttner, Huxford, Huynh-Dinh, Ide,
  Idzkowski, Iess, Ikenoue, Imam, Inayoshi, Ingram, Inoue, Ioka, Isi, Isleif,
  Ito, Itoh, Iyer, Izumi, Jaberian~Hamedan, Jacqmin, Jadhav, Jadhav, James,
  Jan, Jani, Janquart, Janssens, Janthalur, Jaranowski, Jariwala, Jaume,
  Jenkins, Jenner, Jeon, Jeunon, Jia, Jin, Johns, Jones, Jones, Jones, Jones,
  Jones, Jonker, Ju, Jung, Jung, Junker, Juste, Kaihotsu, Kajita, Kakizaki,
  Kalaghatgi, Kalogera, Kamai, Kamiizumi, Kanda, Kandhasamy, Kang, Kanner, Kao,
  Kapadia, Kapasi, Karat, Karathanasis, Karki, Kashyap, Kasprzack, Kastaun,
  Katsanevas, Katsavounidis, Katzman, Kaur, Kawabe, Kawaguchi, Kawai, Kawasaki,
  Kéfélian, Keitel, Key, Khadka, Khalili, Khan, Khazanov, Khetan, Khursheed,
  Kijbunchoo, Kim, Kim, Kim, Kim, Kim, Kim, Kimball, Kimura, Kinley-Hanlon,
  Kirchhoff, Kissel, Kita, Kitazawa, Kleybolte, Klimenko, Knee, Knowles,
  Knyazev, Koch, Koekoek, Kojima, Kokeyama, Koley, Kolitsidou, Kolstein,
  Komori, Kondrashov, Kong, Kontos, Koper, Korobko, Kotake, Kovalam, Kozak,
  Kozakai, Kozu, Kringel, Krishnendu, Królak, Kuehn, Kuei, Kuijer, Kumar,
  Kumar, Kumar, Kumar, Kume, Kuns, Kuo, Kuo, Kuromiya, Kuroyanagi, Kusayanagi,
  Kuwahara, Kwak, Lagabbe, Laghi, Lalande, Lam, Lamberts, Landry, Lane, Lang,
  Lange, Lantz, La~Rosa, Lartaux-Vollard, Lasky, Laxen, Lazzarini, Lazzaro,
  Leaci, Leavey, Lecoeuche, Lee, Lee, Lee, Lee, Lee, Lee, Lehmann, Lemaître,
  Leonardi, Leroy, Letendre, Levesque, Levin, Leviton, Leyde, Li, Li, Li, Li,
  Li, Li, Lin, Lin, Lin, Lin, Lin, Linde, Linker, Linley, Littenberg, Liu, Liu,
  Liu, Liu, Llamas, Llorens-Monteagudo, Lo, Lockwood, London, Longo, Lopez,
  Lopez~Portilla, Lorenzini, Loriette, Lormand, Losurdo, Lott, Lough, Lousto,
  Lovelace, Lucaccioni, Lück, Lumaca, Lundgren, Luo, Lynam, Macas, MacInnis,
  Macleod, MacMillan, Macquet, Magaña~Hernandez, Magazzù, Magee, Maggiore,
  Magnozzi, Mahesh, Majorana, Makarem, Maksimovic, Maliakal, Malik, Man,
  Mandic, Mangano, Mango, Mansell, Manske, Mantovani, Mapelli, Marchesoni,
  Marchio, Marion, Mark, Márka, Márka, Markakis, Markosyan, Markowitz, Maros,
  Marquina, Marsat, Martelli, Martin, Martin, Martinez, Martinez, Martinez,
  Martinovic, Martynov, Marx, Masalehdan, Mason, Massera, Masserot, Massinger,
  Masso-Reid, Mastrogiovanni, Matas, Mateu-Lucena, Matichard, Matiushechkina,
  Mavalvala, McCann, McCarthy, McClelland, McClincy, McCormick, McCuller,
  McGhee, McGuire, McIsaac, McIver, McRae, McWilliams, Meacher, Mehmet, Mehta,
  Meijer, Melatos, Melchor, Mendell, Menendez-Vazquez, Menoni, Mercer, Mereni,
  Merfeld, Merilh, Merritt, Merzougui, Meshkov, Messenger, Messick, Meyers,
  Meylahn, Mhaske, Miani, Miao, Michaloliakos, Michel, Michimura, Middleton,
  Milano, Miller, Miller, Miller, Millhouse, Mills, Milotti, Minazzoli,
  Minenkov, Mio, Mir, Miravet-Tenés, Mishra, Mishra, Mistry, Mitra,
  Mitrofanov, Mitselmakher, Mittleman, Miyakawa, Miyamoto, Miyazaki, Miyo,
  Miyoki, Mo, Moguel, Mogushi, Mohapatra, Mohite, Molina, Molina-Ruiz, Mondin,
  Montani, Moore, Moraru, Morawski, More, Moreno, Moreno, Mori, Morisaki,
  Moriwaki, Mours, Mow-Lowry, Mozzon, Muciaccia, Mukherjee, Mukherjee,
  Mukherjee, Mukherjee, Mukherjee, Mukund, Mullavey, Munch, Muñiz, Murray,
  Musenich, Muusse, Nadji, Nagano, Nagano, Nagar, Nakamura, Nakano, Nakano,
  Nakashima, Nakayama, Napolano, Nardecchia, Narikawa, Naticchioni, Nayak,
  Nayak, Negishi, Neil, Neilson, Nelemans, Nelson, Nery, Neubauer, Neunzert,
  Ng, Ng, Nguyen, Nguyen, Nguyen, Nguyen~Quynh, Ni, Nichols, Nishizawa,
  Nissanke, Nitoglia, Nocera, Norman, North, Nozaki, Nuttall, Oberling,
  O’Brien, Obuchi, O’Dell, Oelker, Ogaki, Oganesyan, Oh, Oh, Oh, Ohashi,
  Ohishi, Ohkawa, Ohme, Ohta, Okada, Okutani, Okutomi, Olivetto, Oohara, Ooi,
  Oram, O’Reilly, Ormiston, Ormsby, Ortega, O’Shaughnessy, O’Shea,
  Oshino, Ossokine, Osthelder, Otabe, Ottaway, Overmier, Pace, Pagano, Page,
  Pagliaroli, Pai, Pai, Palamos, Palashov, Palomba, Pan, Pan, Panda, Pang,
  Pang, Pankow, Pannarale, Pant, Panther, Paoletti, Paoli, Paolone, Parisi,
  Park, Park, Parker, Pascucci, Pasqualetti, Passaquieti, Passuello, Patel,
  Pathak, Patricelli, Patron, Patrone, Paul, Payne, Pedraza, Pegoraro, Pele,
  Peña~Arellano, Penn, Perego, Pereira, Pereira, Perez, Périgois, Perkins,
  Perreca, Perriès, Petermann, Petterson, Pfeiffer, Pham, Phukon, Piccinni,
  Pichot, Piendibene, Piergiovanni, Pierini, Pierro, Pillant, Pillas, Pilo,
  Pinard, Pinto, Pinto, Piotrzkowski, Pirello, Pitkin, Placidi, Planas,
  Plastino, Pluchar, Poggiani, Polini, Pong, Ponrathnam, Popolizio, Porter,
  Poulton, Powell, Pracchia, Pradier, Prajapati, Prasai, Prasanna, Pratten,
  Principe, Prodi, Prokhorov, Prosposito, Prudenzi, Puecher, Punturo, Puosi,
  Puppo, Pürrer, Qi, Quetschke, Quitzow-James, Raab, Raaijmakers, Radkins,
  Radulesco, Raffai, Rail, Raja, Rajan, Ramirez, Ramirez, Ramos-Buades, Rana,
  Rapagnani, Rapol, Ray, Raymond, Raza, Razzano, Read, Rees, Regimbau, Rei,
  Reid, Reid, Reitze, Relton, Renzini, Rettegno, Rezac, Ricci, Richards,
  Richardson, Richardson, Riemenschneider, Riles, Rinaldi, Rink, Rizzo,
  Robertson, Robie, Robinet, Rocchi, Rodriguez, Rolland, Rollins, Romanelli,
  Romano, Romel, Romero-Rodríguez, Romero-Shaw, Romie, Ronchini, Rosa, Rose,
  Rosińska, Ross, Rowan, Rowlinson, Roy, Roy, Roy, Rozza, Ruggi, Ryan,
  Sachdev, Sadecki, Sadiq, Sago, Saito, Saito, Sakai, Sakai, Sakellariadou,
  Sakuno, Salafia, Salconi, Saleem, Salemi, Samajdar, Sanchez, Sanchez,
  Sanchez, Sanchis-Gual, Sanders, Sanuy, Saravanan, Sarin, Sassolas, Satari,
  Sathyaprakash, Sato, Sato, Sauter, Savage, Sawada, Sawant, Sawant, Sayah,
  Schaetzl, Scheel, Scheuer, Schiworski, Schmidt, Schmidt, Schnabel,
  Schneewind, Schofield, Schönbeck, Schulte, Schutz, Schwartz, Scott, Scott,
  Seglar-Arroyo, Sekiguchi, Sekiguchi, Sellers, Sengupta, Sentenac, Seo,
  Sequino, Sergeev, Setyawati, Shaffer, Shahriar, Shams, Shao, Sharma, Sharma,
  Shawhan, Shcheblanov, Shibagaki, Shikauchi, Shimizu, Shimoda, Shimode,
  Shinkai, Shishido, Shoda, Shoemaker, Shoemaker, ShyamSundar, Sieniawska,
  Sigg, Singer, Singh, Singh, Singha, Sintes, Sipala, Skliris, Slagmolen,
  Slaven-Blair, Smetana, Smith, Smith, Soldateschi, Somala, Somiya, Son, Soni,
  Soni, Sordini, Sorrentino, Sorrentino, Sotani, Soulard, Souradeep, Sowell,
  Spagnuolo, Spencer, Spera, Srinivasan, Srivastava, Srivastava, Staats,
  Stachie, Steer, Steinlechner, Steinlechner, Stops, Stover, Strain, Strang,
  Stratta, Strunk, Sturani, Stuver, Sudhagar, Sudhir, Sugimoto, Suh,
  Summerscales, Sun, Sun, Sunil, Sur, Suresh, Sutton, Suzuki, Suzuki, Swinkels,
  Szczepańczyk, Szewczyk, Tacca, vTagoshi, Tait, Takahashi, Takahashi,
  Takamori, Takano, Takeda, Takeda, Talbot, Talbot, Tanaka, Tanaka, Tanaka,
  Tanaka, Tanaka, Tanasijczuk, Tanioka, Tanner, Tao, Tao, Tapia San~Martin,
  Tapia San~Martín, Taranto, Tasson, Telada, Tenorio, Terhune, Terkowski,
  Thirugnanasambandam, Thomas, Thomas, Thompson, Thondapu, Thorne, Thrane,
  Tiwari, Tiwari, Tiwari, Toivonen, Toland, Tolley, Tomaru, Tomigami, Tomura,
  Tonelli, Torres-Forné, Torrie, Tosta~e Melo, Töyrä, Trapananti, Travasso,
  Traylor, Trevor, Tringali, Tripathee, Troiano, Trovato, Trozzo, Trudeau,
  Tsai, Tsai, Tsang, Tsang, Tsao, Tse, Tso, Tsubono, Tsuchida, Tsukada, Tsuna,
  Tsutsui, Tsuzuki, Turbang, Turconi, Tuyenbayev, Ubhi, Uchikata, Uchiyama,
  Udall, Ueda, Uehara, Ueno, Ueshima, Unnikrishnan, Uraguchi, Urban, Ushiba,
  Utina, Vahlbruch, Vajente, Vajpeyi, Valdes, Valentini, Valsan, van Bakel, van
  Beuzekom, van~den Brand, Van Den~Broeck, Vander-Hyde, van~der Schaaf, van
  Heijningen, Vanosky, van Putten, van Remortel, Vardaro, Vargas, Varma,
  Vasúth, Vecchio, Vedovato, Veitch, Veitch, Venneberg, Venugopalan, Verkindt,
  Verma, Verma, Veske, Vetrano, Viceré, Vidyant, Viets, Vijaykumar,
  Villa-Ortega, Vinet, Virtuoso, Vitale, Vo, Vocca, von Reis, von Wrangel,
  Vorvick, Vyatchanin, Wade, Wade, Wagner, Walet, Walker, Wallace, Wallace,
  Walsh, Wang, Wang, Wang, Ward, Warner, Was, Washimi, Washington, Watchi,
  Weaver, Webster, Weinert, Weinstein, Weiss, Weller, Wellmann, Wen, Weßels,
  Wette, Whelan, White, Whiting, Whittle, Wilken, Williams, Williams,
  Williamson, Willis, Willke, Wilson, Winkler, Wipf, Wlodarczyk, Woan, Woehler,
  Wofford, Wong, Wu, Wu, Wu, Wu, Wysocki, Xiao, Xu, Yamada, Yamamoto, Yamamoto,
  Yamamoto, Yamamoto, Yamashita, Yamazaki, Yang, Yang, Yang, Yang, Yang, Yap,
  Yeeles, Yelikar, Ying, Yokogawa, Yokoyama, Yokozawa, Yoo, Yoshioka, Yu, Yu,
  Yuzurihara, Zadrożny, Zanolin, Zeidler, Zelenova, Zendri, Zevin, Zhan,
  Zhang, Zhang, Zhang, Zhang, Zhang, Zhao, Zhao, Zhao, Zhao, Zhou, Zhou, Zhu,
  Zhu, Zimmerman, Zucker, \& Zweizig}]{Abbott_2022}
Abbott, R., Abbott, T.~D., Acernese, F., {et~al.} 2022, Astronomy \&
  Astrophysics, 659, A84.
\newblock \url{http://dx.doi.org/10.1051/0004-6361/202141452}

\bibitem[{Acernese {et~al.}(2014)Acernese, Agathos, Agatsuma, Aisa, Allemandou,
  Allocca, Amarni, Astone, Balestri, Ballardin, \& et~al.}]{AdvVirgo}
Acernese, F., Agathos, M., Agatsuma, K., {et~al.} 2014, Classical and Quantum
  Gravity, 32, 024001.
\newblock \url{http://dx.doi.org/10.1088/0264-9381/32/2/024001}

\bibitem[{Akutsu {et~al.}(2019)Akutsu, Ando, Arai, Arai, Araki, Araya, Aritomi,
  \& et.al.}]{KAGRA}
Akutsu, T., Ando, M., Arai, K., {et~al.} 2019, Nature Astronomy, 3, 35.
\newblock \url{https://doi.org/10.1038%2Fs41550-018-0658-y}

\bibitem[{Apostolatos {et~al.}(1994)Apostolatos, Cutler, Sussman, \&
  Thorne}]{spinprecess}
Apostolatos, T.~A., Cutler, C., Sussman, G.~J., \& Thorne, K.~S. 1994, Phys.
  Rev. D, 49, 6274.
\newblock \url{https://link.aps.org/doi/10.1103/PhysRevD.49.6274}

\bibitem[{{Ara{\'u}jo {\'A}lvarez} {et~al.}(2024){Ara{\'u}jo {\'A}lvarez},
  {Wong}, \& {Calder{\'o}n Bustillo}}]{alvarez2024kicking}
{Ara{\'u}jo {\'A}lvarez}, C., {Wong}, H. W.~Y., \& {Calder{\'o}n Bustillo}, J.
  2024, arXiv e-prints, arXiv:2404.00720

\bibitem[{Biwer {et~al.}(2019)Biwer, Capano, De, Cabero, Brown, Nitz, \&
  Raymond}]{PyCBCInf}
Biwer, C.~M., Capano, C.~D., De, S., {et~al.} 2019, Publications of the
  Astronomical Society of the Pacific, 131, 024503.
\newblock \url{http://dx.doi.org/10.1088/1538-3873/aaef0b}

\bibitem[{Bogdanović {et~al.}(2007)Bogdanović, Reynolds, \&
  Miller}]{Bogdanovi_2007}
Bogdanović, T., Reynolds, C.~S., \& Miller, M.~C. 2007, The Astrophysical
  Journal, 661, L147–L150.
\newblock \url{http://dx.doi.org/10.1086/518769}

\bibitem[{{Bonnor} \& {Rotenberg}(1961)}]{1961RSPSA.265..109B}
{Bonnor}, W.~B., \& {Rotenberg}, M.~A. 1961, Proceedings of the Royal Society
  of London Series A, 265, 109

\bibitem[{Borchers \& Ohme(2023)}]{Borchers_2023}
Borchers, A., \& Ohme, F. 2023, Classical and Quantum Gravity, 40, 095008.
\newblock \url{http://dx.doi.org/10.1088/1361-6382/acc5da}

\bibitem[{Borhanian(2021)}]{Borhanian_2021}
Borhanian, S. 2021, Classical and Quantum Gravity, 38, 175014.
\newblock \url{http://dx.doi.org/10.1088/1361-6382/ac1618}

\bibitem[{Buscicchio {et~al.}(2021)Buscicchio, Klein, Roebber, Moore, Gerosa,
  Finch, \& Vecchio}]{Buscicchio_2021}
Buscicchio, R., Klein, A., Roebber, E., {et~al.} 2021, Physical Review D, 104,
  doi:10.1103/physrevd.104.044065.
\newblock \url{http://dx.doi.org/10.1103/PhysRevD.104.044065}

\bibitem[{Bustillo {et~al.}(2022)Bustillo, Leong, \&
  Chandra}]{bustillo2022gw190412}
Bustillo, J.~C., Leong, S. H.~W., \& Chandra, K. 2022, GW190412: measuring a
  black-hole recoil direction through higher-order gravitational-wave modes, ,
  , arXiv:2211.03465

\bibitem[{Calderón~Bustillo {et~al.}(2018)Calderón~Bustillo, Clark, Laguna,
  \& Shoemaker}]{Calder_n_Bustillo_2018}
Calderón~Bustillo, J., Clark, J.~A., Laguna, P., \& Shoemaker, D. 2018,
  Physical Review Letters, 121, doi:10.1103/physrevlett.121.191102.
\newblock \url{http://dx.doi.org/10.1103/PhysRevLett.121.191102}

\bibitem[{Campanelli {et~al.}(2007)Campanelli, Lousto, Zlochower, \&
  Merritt}]{Campanelli_2007}
Campanelli, M., Lousto, C.~O., Zlochower, Y., \& Merritt, D. 2007, Physical
  Review Letters, 98, doi:10.1103/physrevlett.98.231102.
\newblock \url{http://dx.doi.org/10.1103/PhysRevLett.98.231102}

\bibitem[{{Cutler} \& {Flanagan}(1994)}]{1994PhRvD..49.2658C}
{Cutler}, C., \& {Flanagan}, {\'E}.~E. 1994, Phys. Rev. D, 49, 2658

\bibitem[{{Cutler} {et~al.}(2019){Cutler}, {Berti}, {Jani}, {Kovetz},
  {Randall}, {Vitale}, {Wong}, {Holley-Bockelmann}, {Larson}, {Littenberg},
  {McWilliams}, {Mueller}, {Schnittman}, {Shoemaker}, \&
  {Vallisneri}}]{WP-Multiband}
{Cutler}, C., {Berti}, E., {Jani}, K., {et~al.} 2019, arXiv e-prints,
  arXiv:1903.04069

\bibitem[{Damour(2001)}]{PhysRevD.64.124013}
Damour, T. 2001, Phys. Rev. D, 64, 124013.
\newblock \url{http://link.aps.org/doi/10.1103/PhysRevD.64.124013}

\bibitem[{Doctor {et~al.}(2021)Doctor, Farr, \& Holz}]{Doctor_2021}
Doctor, Z., Farr, B., \& Holz, D.~E. 2021, The Astrophysical Journal Letters,
  914, L18.
\newblock \url{http://dx.doi.org/10.3847/2041-8213/ac0334}

\bibitem[{Dunn {et~al.}(2020)Dunn, Holley-Bockelmann, \& Bellovary}]{Dunn_2020}
Dunn, G., Holley-Bockelmann, K., \& Bellovary, J. 2020, The Astrophysical
  Journal, 896, 72.
\newblock \url{http://dx.doi.org/10.3847/1538-4357/ab7cd2}

\bibitem[{Gerosa \& Berti(2019)}]{Gerosa_2019}
Gerosa, D., \& Berti, E. 2019, Physical Review D, 100,
  doi:10.1103/physrevd.100.041301.
\newblock \url{http://dx.doi.org/10.1103/PhysRevD.100.041301}

\bibitem[{Gerosa \& Fishbach(2021)}]{Gerosa_2021}
Gerosa, D., \& Fishbach, M. 2021, Nature Astronomy, 5, 749–760.
\newblock \url{http://dx.doi.org/10.1038/s41550-021-01398-w}

\bibitem[{González {et~al.}(2007)González, Sperhake, Brügmann, Hannam, \&
  Husa}]{Gonz_lez_2007}
González, J.~A., Sperhake, U., Brügmann, B., Hannam, M., \& Husa, S. 2007,
  Physical Review Letters, 98, doi:10.1103/physrevlett.98.091101.
\newblock \url{http://dx.doi.org/10.1103/PhysRevLett.98.091101}

\bibitem[{Graham {et~al.}(2020)Graham, Ford, McKernan, Ross, Stern, Burdge,
  Coughlin, Djorgovski, Drake, Duev, \& et~al.}]{2020ZTF}
Graham, M., Ford, K., McKernan, B., {et~al.} 2020, Physical Review Letters,
  124, doi:10.1103/physrevlett.124.251102.
\newblock \url{http://dx.doi.org/10.1103/PhysRevLett.124.251102}

\bibitem[{Harms {et~al.}(2021)Harms, Ambrosino, Angelini, Braito, Branchesi,
  Brocato, Cappellaro, Coccia, Coughlin, Ceca, Valle, Dionisio, Federico,
  Formisano, Frigeri, Grado, Izzo, Marcelli, Maselli, Olivieri, Pernechele,
  Possenti, Ronchini, Serafinelli, Severgnini, Agostini, Badaracco, Bertolini,
  Betti, Civitani, Collette, Covino, Dall’Osso, D’Avanzo, DeSalvo,
  Giovanni, Focardi, Giunchi, Heijningen, Khetan, Melini, Mitri, Mow-Lowry,
  Naponiello, Noce, Oganesyan, Pace, Paik, Pajewski, Palazzi, Pallavicini,
  Pareschi, Pozzobon, Sharma, Spada, Stanga, Tagliaferri, \&
  Votta}]{Harms_2021}
Harms, J., Ambrosino, F., Angelini, L., {et~al.} 2021, The Astrophysical
  Journal, 910, 1.
\newblock \url{http://dx.doi.org/10.3847/1538-4357/abe5a7}

\bibitem[{Holley‐Bockelmann {et~al.}(2008)Holley‐Bockelmann, Gültekin,
  Shoemaker, \& Yunes}]{Holley_Bockelmann_2008}
Holley‐Bockelmann, K., Gültekin, K., Shoemaker, D., \& Yunes, N. 2008, The
  Astrophysical Journal, 686, 829–837.
\newblock \url{http://dx.doi.org/10.1086/591218}

\bibitem[{Izumi \& Jani(2021)}]{Izumi_2021}
Izumi, K., \& Jani, K. 2021, Detection Landscape in the deci-Hertz
  Gravitational-Wave Spectrum (Springer Singapore), 1–14.
\newblock \url{http://dx.doi.org/10.1007/978-981-15-4702-7_50-1}

\bibitem[{Jani \& Loeb(2021)}]{Jani_2021}
Jani, K., \& Loeb, A. 2021, Journal of Cosmology and Astroparticle Physics,
  2021, 044.
\newblock \url{https://doi.org/10.1088/1475-7516/2021/06/044}

\bibitem[{{Jani} {et~al.}(2020){Jani}, {Shoemaker}, \&
  {Cutler}}]{2020NatAs...4..260J}
{Jani}, K., {Shoemaker}, D., \& {Cutler}, C. 2020, Nature Astronomy, 4, 260

\bibitem[{Kimball {et~al.}(2021)Kimball, Talbot, Berry, Zevin, Thrane,
  Kalogera, Buscicchio, Carney, Dent, Middleton, Payne, Veitch, \&
  Williams}]{Kimball_2021}
Kimball, C., Talbot, C., Berry, C. P.~L., {et~al.} 2021, The Astrophysical
  Journal Letters, 915, L35.
\newblock \url{http://dx.doi.org/10.3847/2041-8213/ac0aef}

\bibitem[{Kimura {et~al.}(2021)Kimura, Murase, \& Bartos}]{Kimura_2021}
Kimura, S.~S., Murase, K., \& Bartos, I. 2021, The Astrophysical Journal, 916,
  111.
\newblock \url{http://dx.doi.org/10.3847/1538-4357/ac0535}

\bibitem[{{{LIGO Scientific Collaboration}}(2015)}]{AdvLIGORef}
{{LIGO Scientific Collaboration}}. 2015, Classical and Quantum Gravity, 32,
  074001

\bibitem[{Lousto \& Zlochower(2008)}]{Lousto_2008}
Lousto, C.~O., \& Zlochower, Y. 2008, Physical Review D, 77,
  doi:10.1103/physrevd.77.044028.
\newblock \url{http://dx.doi.org/10.1103/PhysRevD.77.044028}

\bibitem[{Mahapatra {et~al.}(2024{\natexlab{a}})Mahapatra, Chattopadhyay,
  Gupta, Antonini, Favata, Sathyaprakash, \&
  Arun}]{mahapatra2024reconstructing}
Mahapatra, P., Chattopadhyay, D., Gupta, A., {et~al.} 2024{\natexlab{a}},
  Reconstructing the genealogy of LIGO-Virgo black holes, , , arXiv:2406.06390

\bibitem[{Mahapatra {et~al.}(2024{\natexlab{b}})Mahapatra, Chattopadhyay,
  Gupta, Favata, Sathyaprakash, \& Arun}]{mahapatra2024predictions}
---. 2024{\natexlab{b}}, Predictions of a simple parametric model of
  hierarchical black hole mergers, , , arXiv:2209.05766

\bibitem[{Mahapatra {et~al.}(2023)Mahapatra, Favata, \&
  Arun}]{mahapatra2023testing}
Mahapatra, P., Favata, M., \& Arun, K.~G. 2023, Testing general relativity via
  direct measurement of black hole kicks, , , arXiv:2308.08319

\bibitem[{Mahapatra {et~al.}(2021)Mahapatra, Gupta, Favata, Arun, \&
  Sathyaprakash}]{Mahapatra_2021}
Mahapatra, P., Gupta, A., Favata, M., Arun, K.~G., \& Sathyaprakash, B.~S.
  2021, The Astrophysical Journal Letters, 918, L31.
\newblock \url{http://dx.doi.org/10.3847/2041-8213/ac20db}

\bibitem[{McKernan {et~al.}(2019)McKernan, Ford, Bartos, Graham, Lyra, Marka,
  Marka, Ross, Stern, \& Yang}]{McKernan_2019}
McKernan, B., Ford, K. E.~S., Bartos, I., {et~al.} 2019, The Astrophysical
  Journal Letters, 884, L50.
\newblock \url{http://dx.doi.org/10.3847/2041-8213/ab4886}

\bibitem[{Merritt {et~al.}(2004)Merritt, Milosavljevi, Favata, Hughes, \&
  Holz}]{Merritt_2004}
Merritt, D., Milosavljevi, M., Favata, M., Hughes, S.~A., \& Holz, D.~E. 2004,
  The Astrophysical Journal, 607, L9–L12.
\newblock \url{http://dx.doi.org/10.1086/421551}

\bibitem[{{Micic} {et~al.}(2006){Micic}, {Abel}, \&
  {Sigurdsson}}]{2006MNRAS.372.1540M}
{Micic}, M., {Abel}, T., \& {Sigurdsson}, S. 2006, \mnras, 372, 1540

\bibitem[{Pekowsky {et~al.}(2013)Pekowsky, Healy, Shoemaker, \&
  Laguna}]{Pekowsky:2012sr}
Pekowsky, L., Healy, J., Shoemaker, D., \& Laguna, P. 2013, Phys. Rev., D87,
  084008

\bibitem[{Peres(1962)}]{PhysRev.128.2471}
Peres, A. 1962, Phys. Rev., 128, 2471.
\newblock \url{https://link.aps.org/doi/10.1103/PhysRev.128.2471}

\bibitem[{{Planck Collaboration} {et~al.}(2018){Planck Collaboration},
  {Aghanim}, {Akrami}, {Ashdown}, {Aumont}, {Baccigalupi}, {Ballardini},
  {Banday}, {Barreiro}, {Bartolo}, {Basak}, {Battye}, {Benabed}, {Bernard},
  {Bersanelli}, {Bielewicz}, {Bock}, {Bond}, {Borrill}, {Bouchet}, {Boulanger},
  {Bucher}, {Burigana}, {Butler}, {Calabrese}, {Cardoso}, {Carron},
  {Challinor}, {Chiang}, {Chluba}, {Colombo}, {Combet}, {Contreras}, {Crill},
  {Cuttaia}, {de Bernardis}, {de Zotti}, {Delabrouille}, {Delouis}, {Di
  Valentino}, {Diego}, {Dor{\'e}}, {Douspis}, {Ducout}, {Dupac}, {Dusini},
  {Efstathiou}, {Elsner}, {En{\ss}lin}, {Eriksen}, {Fantaye}, {Farhang},
  {Fergusson}, {Fernandez-Cobos}, {Finelli}, {Forastieri}, {Frailis},
  {Franceschi}, {Frolov}, {Galeotta}, {Galli}, {Ganga}, {G{\'e}nova-Santos},
  {Gerbino}, {Ghosh}, {Gonz{\'a}lez-Nuevo}, {G{\'o}rski}, {Gratton},
  {Gruppuso}, {Gudmundsson}, {Hamann}, {Hand ley}, {Herranz}, {Hivon}, {Huang},
  {Jaffe}, {Jones}, {Karakci}, {Keih{\"a}nen}, {Keskitalo}, {Kiiveri}, {Kim},
  {Kisner}, {Knox}, {Krachmalnicoff}, {Kunz}, {Kurki-Suonio}, {Lagache},
  {Lamarre}, {Lasenby}, {Lattanzi}, {Lawrence}, {Le Jeune}, {Lemos},
  {Lesgourgues}, {Levrier}, {Lewis}, {Liguori}, {Lilje}, {Lilley}, {Lindholm},
  {L{\'o}pez-Caniego}, {Lubin}, {Ma}, {Mac{\'\i}as-P{\'e}rez}, {Maggio},
  {Maino}, {Mandolesi}, {Mangilli}, {Marcos-Caballero}, {Maris}, {Martin},
  {Martinelli}, {Mart{\'\i}nez-Gonz{\'a}lez}, {Matarrese}, {Mauri}, {McEwen},
  {Meinhold}, {Melchiorri}, {Mennella}, {Migliaccio}, {Millea}, {Mitra},
  {Miville-Desch{\^e}nes}, {Molinari}, {Montier}, {Morgante}, {Moss}, {Natoli},
  {N{\o}rgaard-Nielsen}, {Pagano}, {Paoletti}, {Partridge}, {Patanchon},
  {Peiris}, {Perrotta}, {Pettorino}, {Piacentini}, {Polastri}, {Polenta},
  {Puget}, {Rachen}, {Reinecke}, {Remazeilles}, {Renzi}, {Rocha}, {Rosset},
  {Roudier}, {Rubi{\~n}o-Mart{\'\i}n}, {Ruiz-Granados}, {Salvati}, {Sandri},
  {Savelainen}, {Scott}, {Shellard}, {Sirignano}, {Sirri}, {Spencer},
  {Sunyaev}, {Suur-Uski}, {Tauber}, {Tavagnacco}, {Tenti}, {Toffolatti},
  {Tomasi}, {Trombetti}, {Valenziano}, {Valiviita}, {Van Tent}, {Vibert},
  {Vielva}, {Villa}, {Vittorio}, {Wand elt}, {Wehus}, {White}, {White},
  {Zacchei}, \& {Zonca}}]{2018arXiv180706209P}
{Planck Collaboration}, {Aghanim}, N., {Akrami}, Y., {et~al.} 2018, arXiv
  e-prints, arXiv:1807.06209

\bibitem[{Pratten {et~al.}(2021)Pratten, García-Quirós, Colleoni,
  Ramos-Buades, Estellés, Mateu-Lucena, Jaume, Haney, Keitel, Thompson, \&
  et~al.}]{Pratten_2021}
Pratten, G., García-Quirós, C., Colleoni, M., {et~al.} 2021, Physical Review
  D, 103, doi:10.1103/physrevd.103.104056.
\newblock \url{http://dx.doi.org/10.1103/PhysRevD.103.104056}

\bibitem[{Punturo {et~al.}(2010)Punturo, Abernathy, \& et.al}]{ETPunturo_2010}
Punturo, M., Abernathy, M., \& et.al. 2010, Classical and Quantum Gravity, 27,
  194002.
\newblock \url{https://doi.org/10.1088/0264-9381/27/19/194002}

\bibitem[{Randall \& Xianyu(2021)}]{Randall_2021}
Randall, L., \& Xianyu, Z.-Z. 2021, The Astrophysical Journal, 914, 75.
\newblock \url{https://dx.doi.org/10.3847/1538-4357/abfd2c}

\bibitem[{Reitze {et~al.}(2019)Reitze, Adhikari, Ballmer, Barish, Barsotti,
  Billingsley, Brown, Chen, Coyne, Eisenstein, Evans, Fritschel, Hall,
  Lazzarini, Lovelace, Read, Sathyaprakash, Shoemaker, Smith, Torrie, Vitale,
  Weiss, Wipf, \& Zucker}]{reitze2019cosmic}
Reitze, D., Adhikari, R.~X., Ballmer, S., {et~al.} 2019, Cosmic Explorer: The
  U.S. Contribution to Gravitational-Wave Astronomy beyond LIGO, , ,
  arXiv:1907.04833

\bibitem[{Schmidt {et~al.}(2015)Schmidt, Ohme, \& Hannam}]{chip}
Schmidt, P., Ohme, F., \& Hannam, M. 2015, Physical Review D, 91,
  doi:10.1103/physrevd.91.024043.
\newblock \url{https://doi.org/10.1103%2Fphysrevd.91.024043}

\bibitem[{Schnittman {et~al.}(2008)Schnittman, Buonanno, van Meter, Baker,
  Boggs, Centrella, Kelly, \& McWilliams}]{PhysRevD.77.044031}
Schnittman, J.~D., Buonanno, A., van Meter, J.~R., {et~al.} 2008, Phys. Rev. D,
  77, 044031.
\newblock \url{https://link.aps.org/doi/10.1103/PhysRevD.77.044031}

\bibitem[{Sedda {et~al.}(2020)Sedda, Berry, Jani, Amaro-Seoane, Auclair, Baird,
  Baker, Berti, Breivik, Burrows, Caprini, Chen, Doneva, Ezquiaga, Ford, Katz,
  Kolkowitz, McKernan, Mueller, Nardini, Pikovski, Rajendran, Sesana, Shao,
  Tamanini, Vartanyan, Warburton, Witek, Wong, \& Zevin}]{Sedda_2020}
Sedda, M.~A., Berry, C. P.~L., Jani, K., {et~al.} 2020, Classical and Quantum
  Gravity, 37, 215011.
\newblock \url{http://dx.doi.org/10.1088/1361-6382/abb5c1}

\bibitem[{{Sesana}(2016)}]{2016PhRvL.116w1102S}
{Sesana}, A. 2016, Physical Review Letters, 116, 231102

\bibitem[{Sperhake {et~al.}(2020)Sperhake, Rosca-Mead, Gerosa, \&
  Berti}]{PhysRevD.101.024044}
Sperhake, U., Rosca-Mead, R., Gerosa, D., \& Berti, E. 2020, Phys. Rev. D, 101,
  024044.
\newblock \url{https://link.aps.org/doi/10.1103/PhysRevD.101.024044}

\bibitem[{{Tagawa} {et~al.}(2024){Tagawa}, {Kimura}, {Haiman}, {Perna}, \&
  {Bartos}}]{2024ApJ...966...21T}
{Tagawa}, H., {Kimura}, S.~S., {Haiman}, Z., {Perna}, R., \& {Bartos}, I. 2024,
  \apj, 966, 21

\bibitem[{{Udall} {et~al.}(2020){Udall}, {Jani}, {Lange}, {O'Shaughnessy},
  {Clark}, {Cadonati}, {Shoemaker}, \&
  {Holley-Bockelmann}}]{2020ApJ...900...80U}
{Udall}, R., {Jani}, K., {Lange}, J., {et~al.} 2020, Astrophys. J. Lett., 900,
  80

\bibitem[{Unnikrishnan(2013)}]{UNNIKRISHNAN_2013}
Unnikrishnan, C.~S. 2013, International Journal of Modern Physics D, 22,
  1341010.
\newblock \url{http://dx.doi.org/10.1142/S0218271813410101}

\bibitem[{Varma {et~al.}(2019)Varma, Gerosa, Stein, Hébert, \&
  Zhang}]{Varma_2019}
Varma, V., Gerosa, D., Stein, L.~C., Hébert, F., \& Zhang, H. 2019, Physical
  Review Letters, 122, doi:10.1103/physrevlett.122.011101.
\newblock \url{http://dx.doi.org/10.1103/PhysRevLett.122.011101}

\bibitem[{Varma {et~al.}(2022)Varma, Biscoveanu, Islam, Shaik, Haster, Isi,
  Farr, Field, \& Vitale}]{Varma_2022}
Varma, V., Biscoveanu, S., Islam, T., {et~al.} 2022, Physical Review Letters,
  128, doi:10.1103/physrevlett.128.191102.
\newblock \url{http://dx.doi.org/10.1103/PhysRevLett.128.191102}

\bibitem[{{{Virgo Collaboration}}(2015)}]{AdvVirgoRef}
{{Virgo Collaboration}}. 2015, Classical and Quantum Gravity, 32, 024001

\bibitem[{{Vitale}(2016)}]{2016PhRvL.117e1102V}
{Vitale}, S. 2016, Physical Review Letters, 117, 051102

\end{thebibliography}

\end{document}